\documentclass[11pt]{article}
\usepackage{epsfig}
\usepackage{graphicx}
\pdfoutput=1
\begin{document}
\pagestyle{myheadings}
\thispagestyle{empty}
\normalsize
\title{
{\bf \sc Astrophysical Stochastic Gravitational Wave Background}
\\}
\author{J.A. de Freitas Pacheco\\
Universit\'e de la C\^ote d'Azur - Observatoire de la C\^ote d'Azur\\
06304 Nice Cedex - France\\
keywords: gravitational waves, supernovas, magnetars, compact binaries}
\date {\today}

\maketitle 

\begin{abstract}

The stochastic gravitational wave background produced by supernovas, magnetars and merger of binaries
constituted by a pair of compact objects is reviewed and updated. The merger of systems composed
by two black holes dominates by far the background signal, whose amplitude in the range
$10-100~Hz$ is above the sensitivity
of the planned Einstein laser interferometer (ET). The background signal at $25~Hz$ estimated
by the LIGO-VIRGO collaboration, based on the available merger detection data, is in good agreement with
the present theoretical predictions.

\end{abstract}

\maketitle 

\section{Introduction}

The astrophysical stochastic	gravitational wave background is the consequence of the superposition of a large
number of unresolved sources formed along the history of the universe. The gravitational wave spectrum of such 
a background contains information about the origin and evolution of the sources, the history of the cosmic star 
formation and the evolution of the initial mass function (IMF) of the progenitors. Prediction of such a spectrum
permits also the identification of the best frequency windows where searches for a relic cosmological 
stochastic background should be done. 

One of the characteristics of the astrophysical background is the so-called ``duty cycle'' {\it D}, which is 
defined by the ratio between the typical duration $\Delta\tau$ of a single burst and the average time between 
two events $\tau_s$. Introducing the frequency of events $f_s = 1/\tau_s$, the duty cycle condition 
can be written as
\begin{equation}
\label{dutycycle} 
D = \int_0^{z_*}(1+z)~\Delta\tau~f_s(z)dz \, 
\end{equation}
where the factor $(1+z)$ was introduced to take into account the time dilation. The upper limit $z_*$ represents the
critical redshift beyond which background becomes truly continuous and fixed by the condition $D \geq 1$. In order
to illustrate this point, in the case where the gravitational signal is originated from the merger of two neutron 
stars, the critical redshift is $z_* \sim 0.25$~\cite{Regimbau2006b}, the precise value depending on the adopted 
cosmic star formation rate (CSFR) and the masses of the components. According to those authors, sources present in 
the redshift interval $0.03 < z < 0.25$ give a duty cycle $D \approx 0.1$, producing a cosmic noise 
dubbed ``pop-corn''. Clearly, for lower redshifts, the sources are expected to be detected individually as it was 
the case of the source $GW170817$ and of the candidates $S190425z$, $S190426c$.

Another important property of the astrophysical  background is its spectrum defined by the 
dimensionless parameter $\Omega_{gw}(\nu_0)$, which is related to the energy flux of gravitational 
waves per frequency interval $F_{\nu_0}$ by the equation
\begin{equation}
\label{spectrum}
\Omega_{gw}(\nu_0) = \frac{1}{\rho_{cr}c^3}\nu_0 F_{\nu_0} \,
\end{equation}
In the equation above $\nu_0$ is the gravitational wave frequency at the observer frame 
and $\rho_{cr} = 3H^2_0/8\pi G$ is the critical matter density required to close the universe. As 
usually, $H_0$ is the present value of the Hubble parameter. 

Possible sources of gravitational waves able to produce a stochastic background are core collapse 
supernovas, rotating neutron stars deformed by strong magnetic fields (magnetars) and the merger of compact binaries 
constituted either by two black holes, two neutron stars or one black hole and one neutron star (see, 
for instance, \cite{Regimbau2011} for a review). Predictions of the gravitational wave  background originated from 
these potential sources depend on two basic aspects: the first is the physical mechanism by the which the 
gravitational radiation is generated and the second is the CSFR that fixes the formation rate of sources. Detection 
methods of such a background were recently revised by~\cite{Cusin2019}.

In the present contribution the gravitational wave background generated by the aforementioned sources will be 
reviewed and updated with respect to previous estimates. 

\section{The cosmic star formation rate} 

Since the CSFR plays a fundamental r\^ole in estimates of the astrophysical background of gravitational
waves, some basic aspects concerning this quantity are here reviewed.

Multi-wavelength surveys performed either with the Hubble or Spitzer space telescopes as well as with different 
large ground-based instruments permitted the discovery of galaxies at redshifts as large 
as $z = 9.1$~\cite{Hashimoto2018}, suggesting that the onset of the star formation activity may have been triggered
about 250 Myr after the Big Bang. The beginning of the star formation process could have occurred even 
earlier ($z \sim 17$) if EDGES observations~\cite{Bowman2018} will be confirmed in the future by independent data.

In general, the star formation rate is estimated from indicators, among others the luminosity of the stellar 
continuum or that of the H$\alpha$ line emission. However all these indicators are affected by dust obscuration and 
corrections may lead to discrepant rates found in the literature for $z > 3-4$ (see, for instance, 
\cite{Madau2014,Kobayashi2013,Hopkins2006}. However, far-infrared data obtained with the Spitzer space telescope 
permit the CSFR to be well determined for redshift $z \leq 1$.

Here the adopted relation describing the CSFR is a compromise between the fit of low-redshift data where dust 
obscuration corrections are more trustful and the fit of high-redshift simulated data by~\cite{Filloux2010,Filloux2011}, which 
explains quite well the photometric properties of galaxies. Such 
a relation as a function of the redshift is given by
\begin{equation}
\label{csfr}
R_*(z) = \frac{(0.0103+0.12z)}{[1+(z/4)^{2.8}]} \,\, M_\odot Mpc^{-3}yr^{-1} \, 
\end{equation}
\begin{figure}
\label{fig:1} 
\centering
 \includegraphics[width=10.00cm]{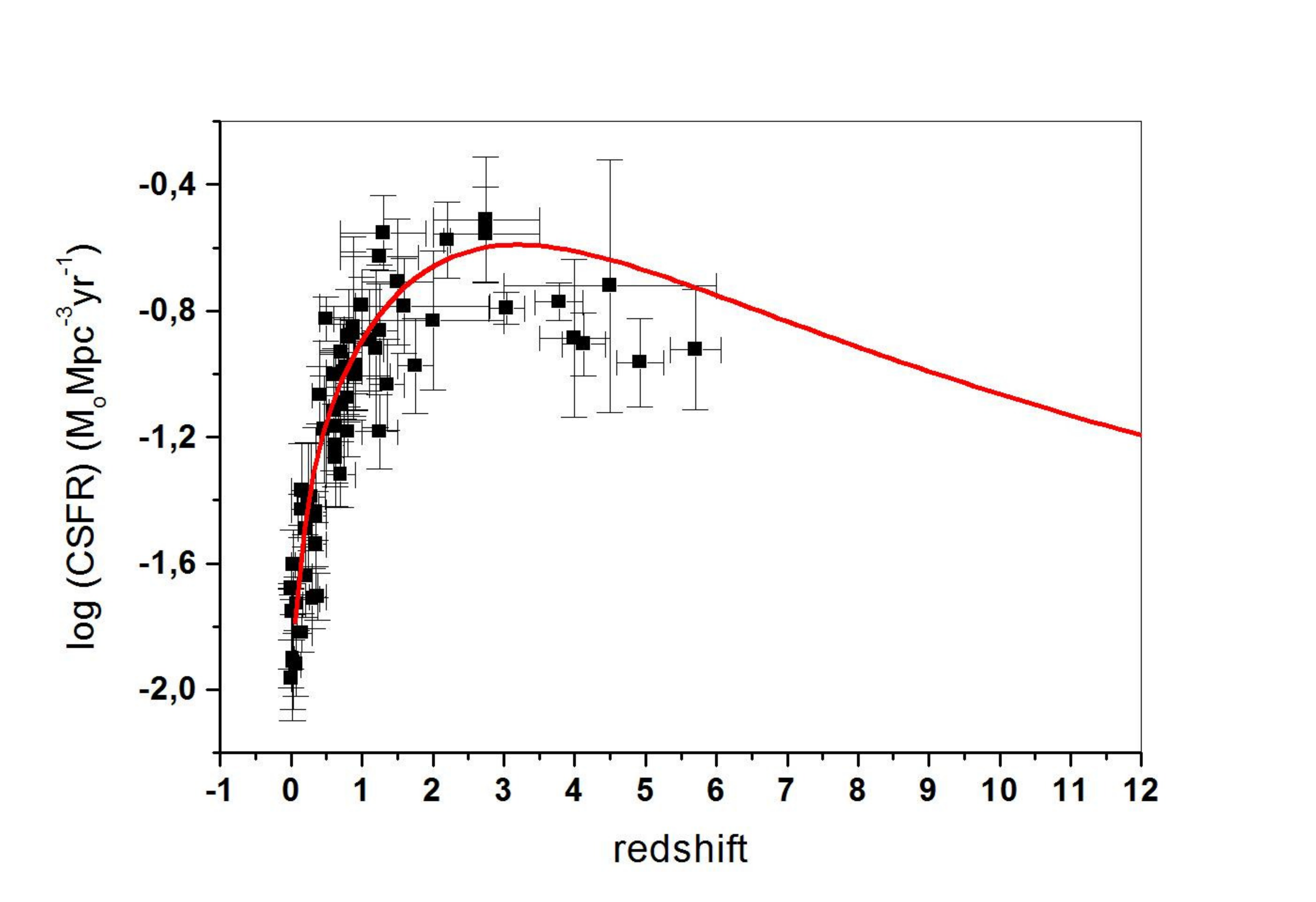}
\caption{Cosmic star formation rate defined by eq.\ref{csfr} (red line) compared with data available in the literature.}
\end{figure}
In figure 1 the adopted fit is compared with data collected from the literature~\cite{Hopkins2004,Hopkins2006}. Notice that 
the fit is quite good for $z \leq 3$ but overestimates the more uncertain data at higher $z$. However such a 
smaller slope in the CSFR for $z > 3$ is necessary to explain observations suggesting the onset of star formation activity 
at $z \sim 9$ (or even higher) as mentioned previously. Moreover, the CSFR modelled by eq.\ref{csfr} describes quite 
well the type II supernova rate per unit of volume as a function of the redshift and the ionization optical depth 
measured by Planck in the CMB.

\section{Gravitational waves from type II supernova}

Stars having masses in the range $9 \leq M/M_\odot \leq 50$ become unstable at the end of their lives, undergo 
the gravitational collapse and explode leaving a neutron star remnant. Gravitational waves are emitted during the 
collapsing phase and during the bounce of the outer layers of the stellar envelope on the hard neutron-rich core 
formed in the very first instants of the process. 

Early investigations on the contribution of supernovas to the astrophysical gravitational wave background 
were performed by~\cite{Ferrari1999}, who concluded that the resulting spectrum is nearly flat in the frequency 
interval $1.5 - 2.5~kHz$, having an amplitude $\Omega_{gw} \sim 10^{-11} - 10^{-12}$.
They concluded also that the expected duty cycle for supernovas is $D < 1$, indicating that the background
is not continuous but being of the ''pop-corn'' type. The contribution of supernovas to the astrophysical gravitational 
background was also considered by~\cite{Buonanno2005}. These authors found a near flat spectrum at frequencies lower 
than those estimated by Ferrari et al. (1999), that is in the interval $100 - 1000~Hz$ with a comparable 
amplitude, i.e., $\Omega_{gw} \sim 10^{-12}$.
\begin{figure}[htb]
\begin{center}
\label{fig:2} 
\centering
\includegraphics[width=8.5cm]{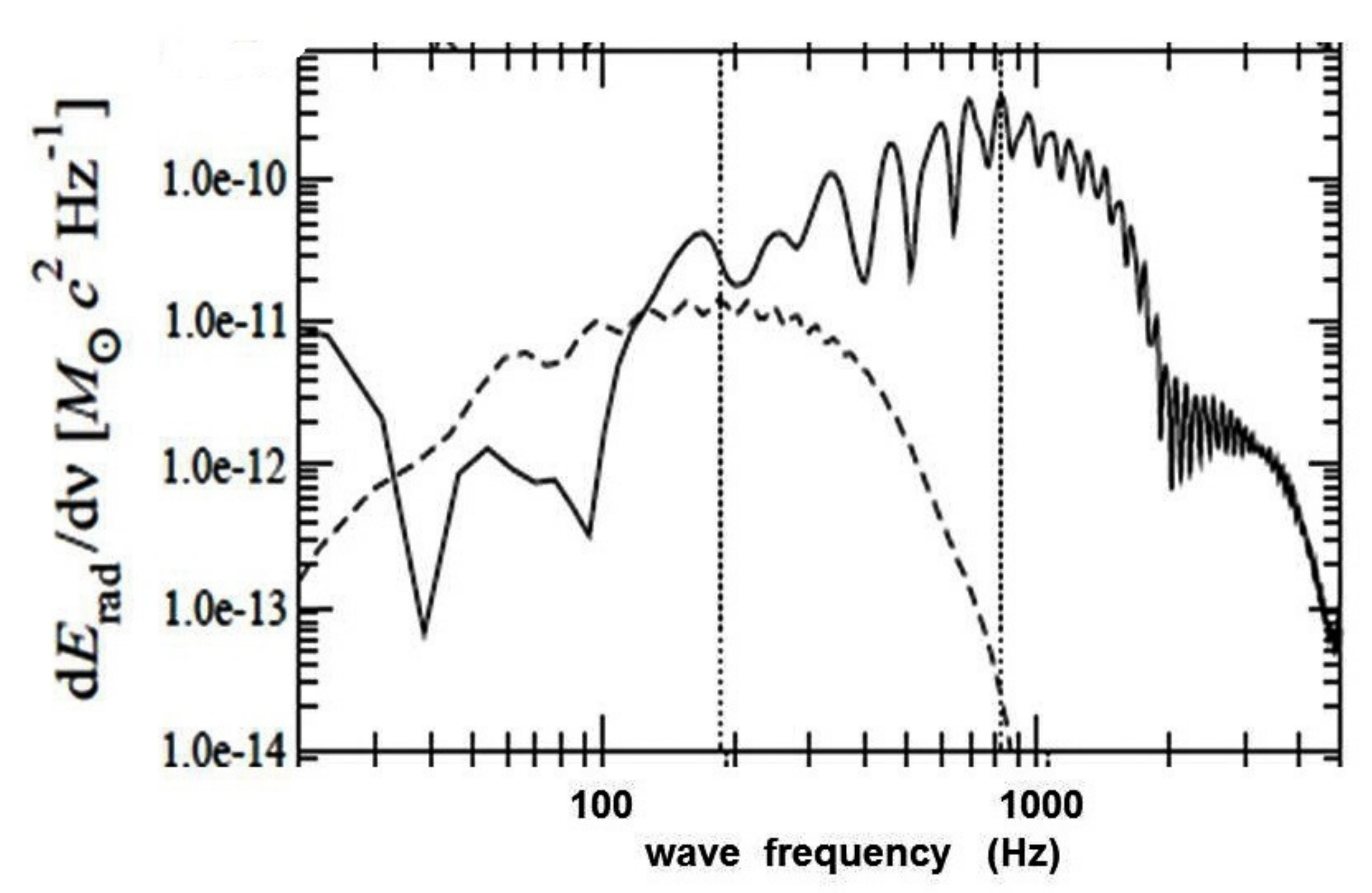}
\caption{Gravitational energy spectrum from core collapse supernova. The dashed line represents
the Newtonian result while the continuous curve represents the relativistic results (from \cite{Dimmelmeier2002})}
\end{center}
\end{figure}
In the past decades a considerable effort has been done to improve our knowledge on the physics describing
the collapse of a star and the associated emission of gravitational radiation. \cite{Dimmelmeier2002} 
performed relativistic calculations of the gravitational collapse of rotating stars and of the 
gravitational waves emitted during the process. They computed a grid of 26 models, assuming that the initial 
star configuration is described by a 4/3-polytropic. The post-bounce phase, in which the outer layers 
of the envelope are ejected by momentum transfer from neutrinos, was considered by~\cite{Murphy2009}, who showed 
that the gravitational wave frequency in this particular phase evolves from about $100~Hz$ up to $300-400~Hz$. 
In reference \cite{Kuroda2016} fully 3D relativistic core collapse computations were performed for a $15~M_\odot$ non-rotating 
star, testing three different equations of state for the nuclear matter. They found that gravitational waves 
with frequencies $\sim 100-200~Hz$ are a typical signature for the standing accretion shock instability (SASI). 
According to~\cite{Andresen2017}, for non-rotating stars, the emission of gravitational waves in the collapsing phase
depends on whether the post-shock is dominated by SASI or convection, since the SASI activity produces
a stronger signal due to asymmetric mass motions. It is worth mentioning that the evolution of the gravitational
wave frequency of the radiation emitted just after the bounce was computed by~\cite{Sotani2016} in the
framework of a relativistic linear perturbation theory. 

In order to compute the background spectrum due to supernovas eq.\ref{spectrum} will be used, taking into
account that the wave frequency $\nu$ in the observer frame and that in the rest frame $\nu_0$ are related
by $\nu = (1+z)\nu_0$. Let $dE/d\nu_0$ be the energy spectrum of gravitational radiation (in the observer
frame) emitted during the collapsing phase and $\lambda_{SN}$ be the fraction by mass of the CSFR that 
will give origin to supernova progenitors. This is equivalent to say, if the lifetime of the progenitors
can be neglected, that the supernova rate per unit co-moving volume at a given redshift is simply given
by $\nu_{SN}(z)=\lambda_{SN}R_*(z)$. Under these conditions, the expected gravitational wave flux at the 
observer frame is
\begin{equation}
\label{flux}
F_{\nu_0} = \lambda_{SN}\int^{z_{max}}_0\frac{1}{4\pi d^2_L}\frac{dE}{d\nu}R_*(z)\frac{dV}{dz}dz
\end{equation}
In the equation above $d_L$ is the luminosity distance, and the element of comoving volume $dV$ is
\begin{equation}
\label{covol}
dV = 4\pi r^2\frac{c}{H_0}\frac{dz}{E(\Omega_i,z)} \, 
\end{equation}
where $r$ is the proper distance, the term $E(\Omega_i,z)$ depends on the cosmology and for the 
standard $\Lambda CDM$ model we have
\begin{equation}
\label{cosmology}
E(\Omega_i,z) = \sqrt{\Omega_V + \Omega_m(1+z)^3} \, 
\end{equation}
where $\Omega_V$ is the equivalent cosmological constant density parameter and $\Omega_m$ is the matter
(dark + baryonic) density parameter.

The value of the parameter $\lambda_{SN}$ can be estimated either from the fit of the expected supernova
rate per unit of volume $\nu_{SN}(z)$ with the available data or from the integral
of the IMF in the mass range $9-50~M_\odot$. Both methods give concordant results
and indicate a value of $\lambda_{SN} = 5.7\times 10^{-3}~M^{-1}_\odot$. For the energy distribution, the
model A3B3G1 by~\cite{Dimmelmeier2002} was adopted. This model is characterized by a 
central density at bounce of $3.5\times 10^{14}~ g.cm^{-3}$ and a bounce timescale of $95~ms$. The 
corresponding energy distribution is shown in figure 2. Replacing the above relation into eq.\ref{spectrum}
one obtains numerically
\begin{equation}
\label{snspectrum}
\Omega_{gw}(\nu_0)h^2 = 3.1\times 10^{-4}\nu_0\int^{z_{max}}_0\frac{R_*(z)}{(1+z)^2}\left(\frac{dE}{d\nu}\right)  
\frac{dz}{E(\Omega_i,z)}
\end{equation}
where $h$ is the Hubble parameter in units of $100~km.s^{-1}.Mpc^{-1}$. The value of $z_{max}$ depends on the 
considered frequency $\nu_0$ and on the maximum frequency of the gravitational waves emitted in 
the process that for the adopted model is about $6~kHz$. Hence,
$z_{max} = (6~kHz/\nu_0)-1$. Figure 3 shows the resulting spectrum computed under the conditions above.
\begin{figure}[htb]
\begin{center}
\label{fig:3} 
\centering
\includegraphics[width=9.50cm]{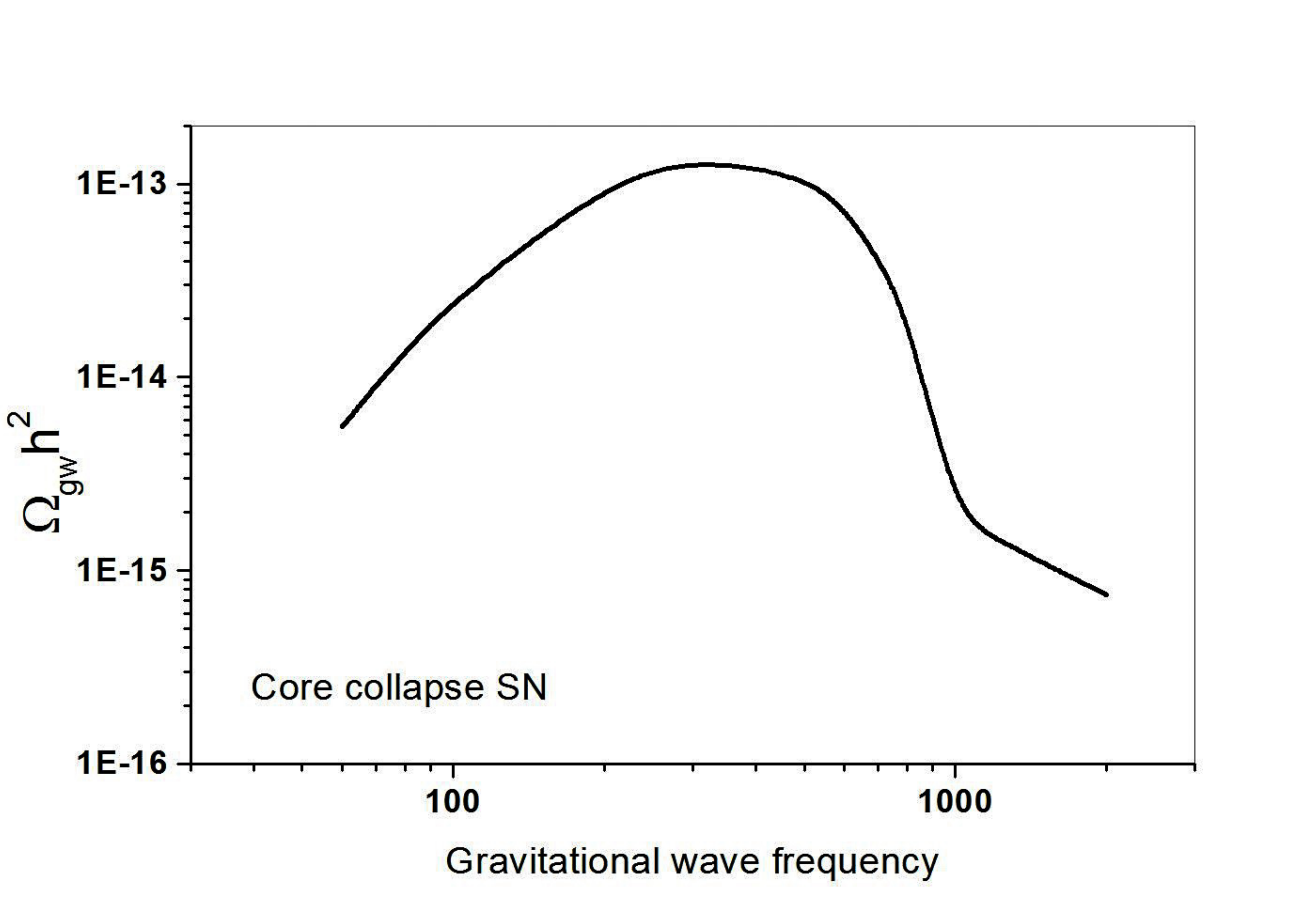}
\caption{Gravitational wave background spectrum due to core collapse supernovas}
\end{center}
\end{figure}
Simple inspection of figure 3 shows that the background spectrum of gravitational waves due to
supernovas has a broad maximum around $300~Hz$, being compatible with the results of~\cite{Buonanno2005}
but with an amplitude one order of magnitude lower, i.e., $\Omega_{gw} \sim 10^{-13}$. 

\section{Gravitational waves from magnetars}

The possibility that rotating tri-axial neutron stars could generate a continuous background was
investigated two decades ago by~\cite{Regimbau2001b}. In their investigation, the
authors assumed that neutron stars have an average ellipticity $\varepsilon = 10^{-6}$ and adopted
CSFR relations quite distinct from eq.\ref{csfr}. They concluded that the resulting
background has a maximum around $0.9-1.5~kHz$ and an amplitude of $\Omega_{gw} \sim 10^{-11} - 10^{-9}$, which
depends not only on the adopted CSFR but also on the adopted maximum rotation frequency of the neutron star
at birth. Later, the same authors considered a sub-population of neutron stars (magnetars) distorted 
by strong magnetic fields~\cite{Regimbau2006a}, hereafter RP06). They found that the resulting 
background spectrum has a broad maximum around $1.2~kHz$ with an amplitude $\Omega_{gw} \sim 10^{-9}$, if
neutron stars are supposed to have a type I superconducting interior.

Here these estimates are revisited with two main differences: the first is the CSFR that is
now given by eq.\ref{csfr} and the second corresponds to the abandon of the superconducting interior
model, what reduces by few orders of magnitude the amplitude of the background signal, since the resulting 
ellipticities are considerably smaller. In order
to compute the background spectrum, the same procedure as before will be adopted. Hence, the energy
spectrum will be first computed as
\begin{equation}
\label{magenergy}
\frac{dE}{d\nu} = \frac{dE}{dt}\mid\frac{dt}{d\nu}\mid \, 
\end{equation}
In the equation above, the term $dE/dt$ represents the quadrupole gravitational energy emission rate by 
the rotating neutron star. However, the dominant mechanism by which the star loses angular momentum is
the magnetic dipole radiation (if the ellipticity is lower than $10^{-4}$). Thus, the evolution of the 
rotation frequency is controlled by this mechanism.
Under these conditions, eq.\ref{magenergy} can be recast as
\begin{equation}
\frac{dE}{d\nu} = K(B)~\nu^3 \, , 
\end{equation}
where the parameter $K(B)$ (function of the magnetic field) is defined as
\begin{equation}
K(B) = \frac{384\pi^4}{5c^2}\frac{GI^3_{zz}}{R^6B^2}\varepsilon^2_B
\end{equation}
where $I_{zz}$ is the moment of inertia along the rotation axis, $R$ is the average neutron star radius,
$B$ is magnetic field at the surface and $\varepsilon_B$ is the ellipticity produced by the internal
magnetic stresses. The other symbols having their usual meaning. If the neutron star is distorted
essentially by magnetic stresses, the ellipticity can be expressed as \cite{Konno2000}
\begin{equation}
\label{ellipticity}
\varepsilon_B = g\frac{R^4B^2sin^2\alpha}{GM^2} \, 
\end{equation}
where $M$ is the mass of the neutron star, $\alpha$ is the angle between the rotation and the 
magnetic dipole axes and $g$ is a dimensionless parameter that depends on the
equation of state and on the internal magnetic field configuration. Here the value $g = 13$ will be
adopted since it is more adequate for non-superconductive neutron star models. It is assumed
that the dipole inclination angle $\alpha$ has a random orientation.
\begin{figure}
\begin{center}
\label{fig:4} 
\includegraphics[width=9.50cm]{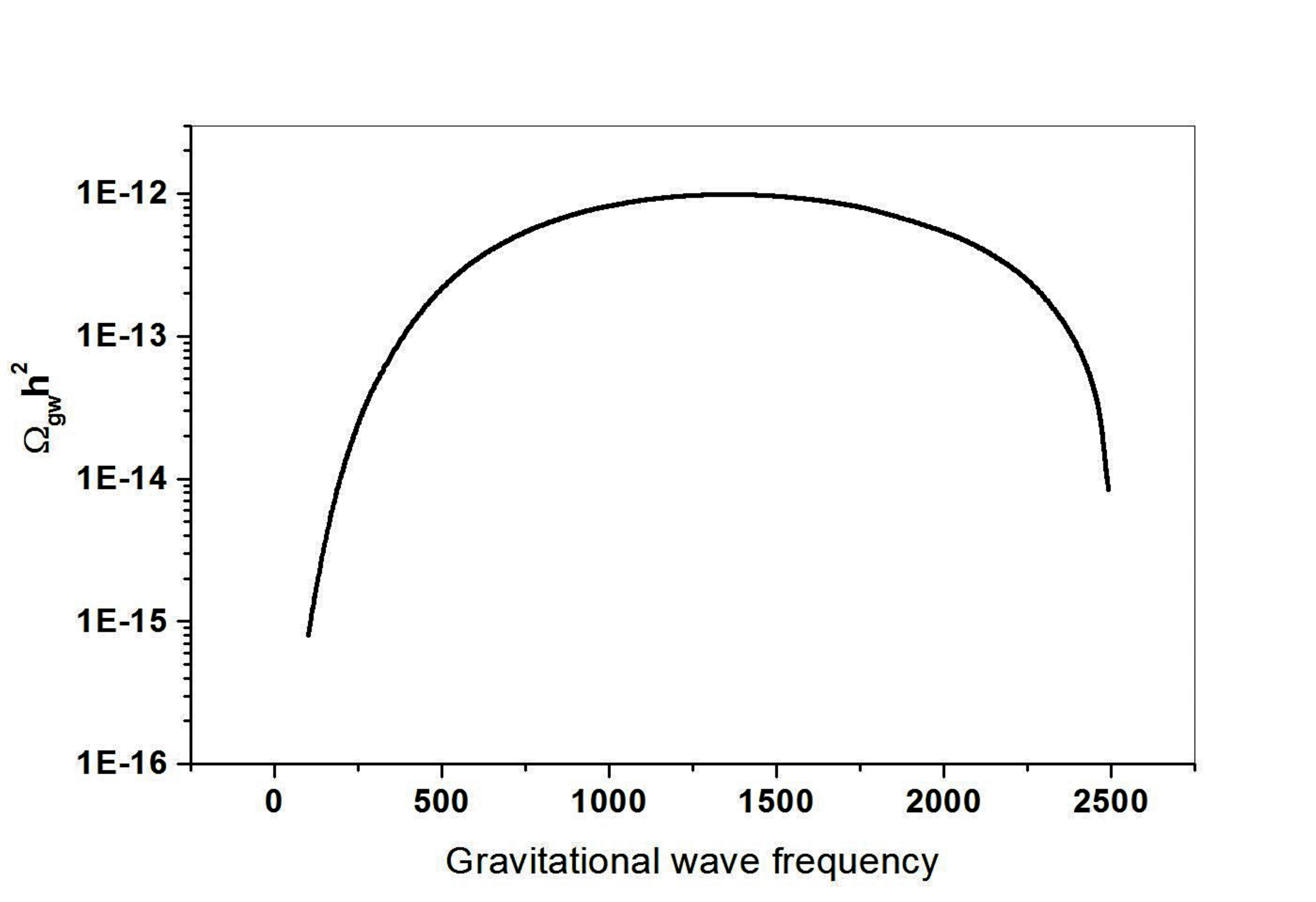}
\caption{Gravitational wave background spectrum due to magnetars}
\end{center}
\end{figure}

We assume also that the formation rate of neutron stars is equal to that of core collapse supernovas but only
a fraction having magnetic fields higher than $10^{14}~G$ are considered ``magnetars". Let $dp(B)/dB$ the normalized
magnetic field  distribution of neutron stars (see RP06). According to the population synthesis simulations
by~\cite{Regimbau2001a}, the true magnetic field distribution of pulsars (not the
observed one) can be represented by a log-normal of mean $log~B = 13.0$ (in Gauss) and with a 
dispersion $\sigma_{log B}=0.8$. Consequently, the formation rate of magnetars per comoving volume and per unit of magnetic
field strength can be written as
\begin{equation}
\label{magrate}
\frac{dR_{mag}}{dB} = \lambda_{SN}\frac{dp(B)}{dB}R_*(z) \, 
\end{equation}
With these definitions, the expected gravitational wave flux results to be
\begin{equation}
\label{fluxmag}
F_{\nu_0}=\lambda_{SN}\frac{c}{H_0}\nu_0^3\int^{z_{max}}_0\frac{(1+z)R_*(z)}{E(\Omega_i,z)}dz\int^{\infty}_{B_c}
K(B)\frac{dp(B)}{dB}dB
\end{equation}
In the one side, the upper limit in the first integral is estimated by taking the minimum rotation 
period of pulsars as being $P_o = 0.8~ms$. Recalling that the gravitational wave frequency is twice the 
rotation frequency, this implies that $z_{max} =(2.5~kHz/\nu_0)-1$. On the other side, the lower limit of
the second integral is taken to be $B_c = 10^{14}~G$, the lower field limit characterizing magnetars. Replacing
eq.\ref{fluxmag} into eq.\ref{spectrum} one obtains
\begin{equation}
\label{magspectrum}
\Omega_{gw}h^2 = 5.12\times 10^{-24}~\nu_0^4\int^{z_{max}}_0\frac{(1+z)R_*(z)}{E(\Omega_i,z)}dz \, 
\end{equation}
In the numerical calculations, a ''canonical'' neutron star model was assumed, i.e., having the
following parameters: $M = 1.4~M_\odot$, $R = 10~km$ and $I_{zz}= 1.4\times 10^{45}~g.cm^2$. The
resulting spectrum computed from eq.\ref{magspectrum} is shown in figure 4. Again, a broad
maximum is seen around $1.3~kHz$ but with an amplitude ($\Omega_{gw} \sim 10^{-12}$) smaller
than that previously estimated by RP06.

\section{Gravitational waves from binary mergers}

The merger of binary systems composed by two compact objects (two neutron stars, two black holes or one neutron
star and one black hole) are among the most important sources of gravitational waves in the universe. However,
predictions of the amplitude of generated background signal are still quite uncertain mainly because 
estimates of merger rates are not robust. An additional difficulty concerns the evolution timescale of the
binary system since its formation up to the beginning of the inspiral phase dominated by the radiation of
gravitational waves. The first or the nuclear fuel phase, depends essentially on the mass of the 
progenitors (including processes of mass loss and mass exchange) and lasts until the formation of a dynamically 
stable binary constituted by two compact objects.
The former phase is characterized by a timescale $t_{min}$ whereas the later or the inspiral phase depends essentially 
on the masses of the compact objects and on their separation at the moment of their formation. 
In order to characterize such a phase, we introduce the probability distribution per unit of time
$P(\tau)$, which measures the fraction of mergers occurring at redshift $z$ in a timescale $\tau$ defined
by the time interval between the instant at which the two compact objects appear and the instant they merge together. Hence, the
merger rate $\rho_{b,i}$ is obtained by the convolution between the probability distribution $P(\tau)$ and the 
CSFR. Merger simulations of binaries constituted by a neutron star pair indicate that $P(\tau) \propto 1/\tau$ (\cite{Pacheco2006}, 
hereafter PRVS06). This probability distribution with be adopted here for the composition of all
pairs. Under these assumptions, the merger rate per unit of comoving volume for a given pair composition $''i''$
is
\begin{equation}
\label{binrate}
\rho_{b,i} = \lambda_{b,i}\int^{\infty}_{z_{c(z)}}R_*(z')P(\tau(z',z))\mid\frac{dt}{dz'}\mid dz' \, 
\end{equation}
where $\lambda_{b,i}$ is the fraction by mass of formed stars producing a pair of massive stars that remain
bounded after exhaustion of their nuclear fuel. The time interval between the formation of the compact objects
at redshift $z'$ and their merger at redshift $z$ is calculated as
\begin{equation}
\tau(z',z) = \frac{1}{H_0}\int^{z'}_z\frac{dx}{(1+x)E(\Omega_i,x)} \, 
\end{equation}
The lower limit $z_c(z)$ of the integral in eq.\ref{binrate} is related to the average timescale of the nuclear
evolution of the progenitors and should be calculated from the integral equation
\begin{equation}  
t_{min}H_0 =\int^{z_c(z)}_z\frac{dx}{(1+x)E(\Omega_i,x)} \, . 
\end{equation}

In the case of a binary constituted by two neutron stars, the parameter $\lambda_{b,NSNS}$ is estimated
from the value of the local ($z = 0$) merger rate, that is
$612~Gpc^{-3}.yr^{-1}$ (PRVS06). This value agrees quite well with a recent independent estimate 
by~\cite{Mapelli2018}, who derived a local NS-NS merger rate of $591~Gpc^{-3}.yr^{-1}$. For binaries
having a different pair composition, relative rates based on population synthesis ~\cite{Belczynski2006} or on 
cosmological simulations were adopted~\cite{Mapelli2018}. The final
relative rates used in the present calculations were: $(NS-BH)/(NS-NS) = 0.12$ and $(BH-BH)/(NS-NS) = 0.19$.

The maximum gravitational wave frequency corresponds approximately to twice the orbital frequency of the
last stable orbit and it can be estimated from the expression~\cite{Buonanno1999}
\begin{equation}
\nu_{max} = 4397(1+0.316\eta)\left[\frac{M_\odot}{(M_1 + M_2)}\right] \,\, Hz \,  
\end{equation}
where $M_1$ and $M_2$ are the masses of the components of the pair in solar units and the parameter $\eta$
is defined by $\eta = \mu/M$ or, in other words, by the ratio between the reduced mass $\mu$ and the 
total mass $M$ of the system. For neutron star pairs, both components were assumed to have masses
equal to $1.4~M_\odot$. Consequently, the maximum emitted frequency is about 1690 Hz. For systems having a
black hole, simulations were performed by assuming that the probability for the progenitor to have a mass $M_*$
is given by IMF and that the minimal mass of the progenitor is $50~M_\odot$. The black
hole mass $M_{bh}$ relates to that of the progenitor by $M_{bh} =\alpha M_*$ with $\alpha = 0.2$. At higher
$z$ or at lower metallicities the IMF becomes more flatter favouring the formation of more massive stars and
more massive black holes. The median of the resulting maximum frequency distributions derived from these
simulations are $300~Hz$ for binaries having one neutron star and one black hole and $170~Hz$ for binaries
constituted by two black holes.
\begin{figure}
\begin{center}
\label{fig:5} 
\includegraphics[width=9.50cm]{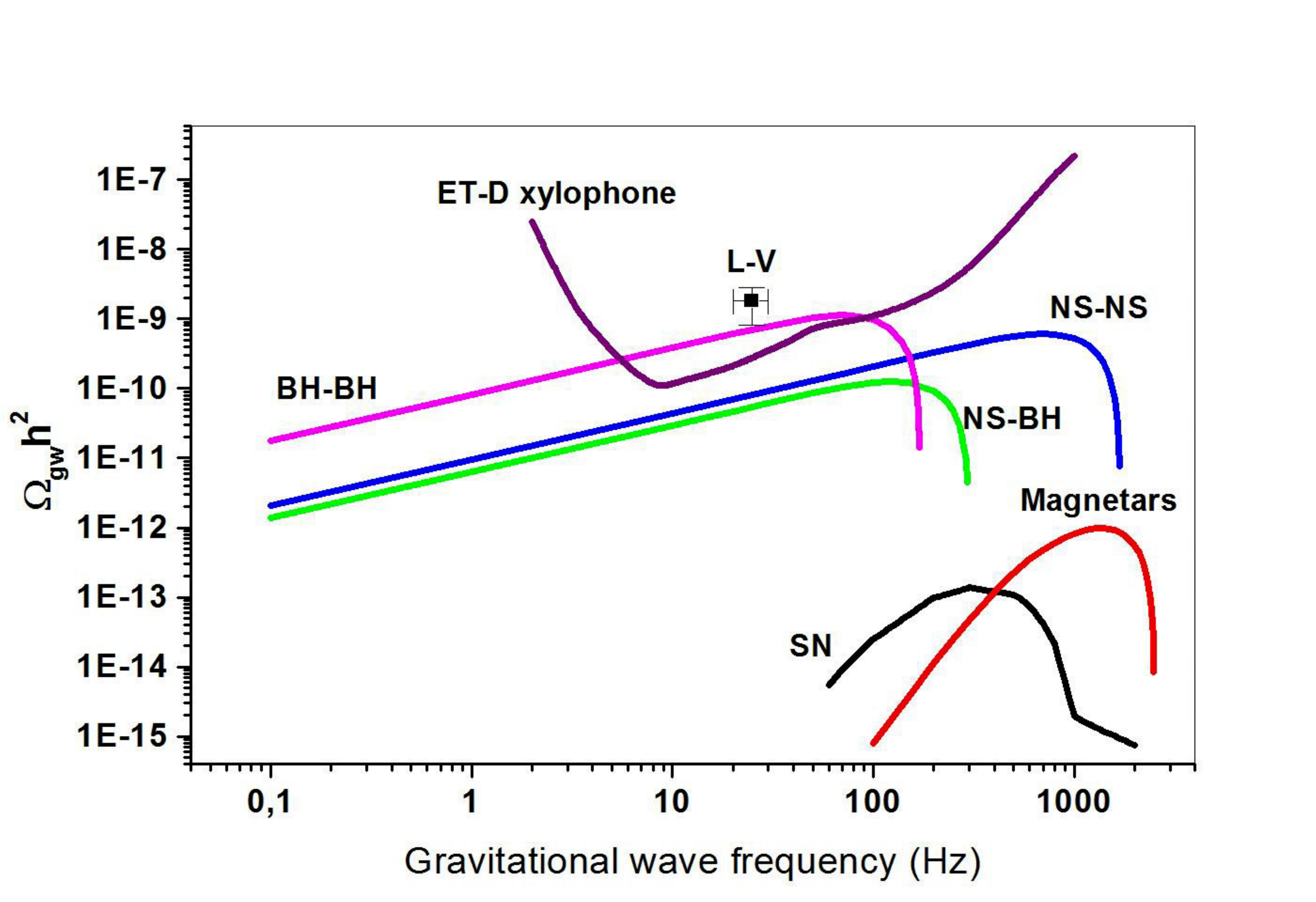}
\caption{Gravitational wave background spectrum due to the merger of compact pairs: BH-BH (pink line),
NS-BH (green line) and NS-NS (blue line). For comparison, the spectrum of supernovas and magnetars
are also plotted. The sensitivity curve of the planned ET laser antenna in its D-xylophone version is
also shown. The gravitational background amplitude at $25~Hz$ estimated by the LIGO-VIRGO team is equally
shown in the plot}
\end{center}
\end{figure}

The last point to be considered is the energy spectrum of the emitted waves. In the quadrupolar
approximation, assuming circular orbits one obtains for energy spectrum  
\begin{equation}
\frac{dE}{d\nu} = K_m~\nu^{-1/3} \,  
\end{equation}
where the parameter $K_m$ is defined by
\begin{equation}
K_m = \frac{(G\pi)^{2/3}}{3}\frac{M_1M_2}{(M_1+M_2)^{1/3}}  \, 
\end{equation}
Similar simulations as those used to estimate maximum frequencies were used in order to compute the median
values of $K_m$. These are respectively for NS-NS, NS-BH and BH-BH pairs, 
$5.2 \times 10^{50}$, $2.9\times 10^{51}$ and $2.3\times 10^{52}$ (in $erg.Hz^{-2/3}$).

From the relations above, the expected flux at the observer frame is
\begin{equation}
\label{fluxbin}
F_{\nu_0,i} = \frac{K_{m,i}}{H_0}\frac{c}{\nu_0^{1/3}}\int^{z_{max}}_0\frac{\rho_{b,i}}{(1+z)^{7/3}}\frac{dz}{E(\Omega_i,z)} \, 
\end{equation}
where again, for each pair composition $z_{max}=(\nu_{max,i}/\nu_0)-1)$.

Finally, the spectrum for each binary composition was computed from eq.\ref{spectrum}, using the flux given
by eq.\ref{fluxbin}. The results are plotted in figure 5 together with the spectra of supernovas and magnetars
for comparison. As expected, the most important astrophysical background signal is due to the merger of BH-BH pairs 
that has a maximum around $100~Hz$ with an amplitude of $\Omega_{gw} \sim 10^{-9}$ while the merger of NS-BH pairs
produce a signal with a maximum near $150~Hz$ and an amplitude $25$ times lower. NS-NS mergers produce a background
signal whose maximum is about $900~Hz$ and the amplitude is $\Omega_{gw} \sim 3\times 10^{-10}$. This should be
compared with the investigation by~\cite{Regimbau2007}, who found a maximum around $500~Hz$ with
an amplitude of $\Omega_{gw} \sim 7\times 10^{-10}$.

\section{Conclusions}

In this work the astrophysical stochastic background of gravitational waves was revisited with emphasis on
the contribution of supernovas, magnetars and merger of binaries constituted by compact objects. Special care
was taken in the choice of the CSFR that combines data from lower redshifts with simulated data at high $z$.

The revised contribution of supernovas and magnetars indicates amplitudes lower than previous estimates and
considerably smaller than the signal produced by the merger of binaries including neutron stars 
and/or black holes, which are the most important sources of the astrophysical stochastic background.

It should be emphasized that the planned gravitational wave telescope Einstein in its D-xylophone version
has a sensitivity adequate to detect the background signal originated from mergers of binaries and in particular
from that produced by BH-BH pairs, the dominant component. Using the existent information 
based on detected merger events, the LIGO-VIRGO
collaboration~\cite{Abbott2018} has estimated an amplitude of $\Omega_{gw} \sim 10^{-9}$ at $25~Hz$, which
is quite compatible with the theoretical estimates here presented.


\begin{thebibliography}{}
\bibitem{Abbott2018}
Abbott, B. et al. 2018, (LIGO-VIRGO Scientific Collaboration) Phys. Rev. Lett., 120, 091101.
\bibitem{Andresen2017}
Andresen, H., Muller, B., Muller, E. \& Janka, H.-T. 2017, Mon.Not.R.Astron.Soc., 468, 2032.
\bibitem{Belczynski2006}
Belczynski, K., Taam, R., Kalogera, V., Rasio, F. \& Bulik, T. 2006, Astrophys.J., 662, 504.
\bibitem{Bowman2018}
Bowman, J., Rogers, A., Monsalve, R., Mozden, T. \& Mahesh, N. 2018, Nature, 555, 67.
\bibitem{Buonanno1999}
Buonanno, A., \& Damour, T. 1999, Phys.Rev. D, 59, 084006.
\bibitem{Buonanno2005}
Buonanno, A., Sigla, G., Raffelt, G., Janka, H.-T. \& Muller, E. 2005, Phys.Rev. D, 72, 084001.
\bibitem{Cusin2019}
Cusin, G., Dvorkin, I., Pitrou, C. \& Uzan, J.-P. 2019, arXiv:astro-ph/1904.07797.
\bibitem{Pacheco2006}
de Freitas Pacheco, J., Regimbau, T., Vincent, S. \& Spallici, A. 2006, (PRVS06), Int. J. Mod. Phys. D, 15, 235.
\bibitem{Dimmelmeier2002}
Dimmelmeier, H., Font, J. \& Muller, E. 2002, Astron.\& Astrophys., 393, 523.
\bibitem{Ferrari1999}
Ferrari, V., Matarrese, S. \& Schneider, R. 1999, Mon. Not. R. Astron. Soc., 303, 247.
\bibitem{Filloux2011}
Filloux, C., de Freitas Pacheco, J., Durier, F. \& de Araújo, J.C.N. 2011, Int. J. Mod. Phys. D, 20, 2399.
\bibitem{Filloux2010}
Filloux, C., Durier, F., de Freitas Pacheco, J. \& Silk, J. 2010, Int. J. Mod. Phys. D, 19, 1233.
\bibitem{Hashimoto2018}
Hashimoto, T., Laporte, N. \& Mawatari, K. et al. 2018, Nature, 557, 392.
\bibitem{Hopkins2004}
Hopkins, A. 2004, Astrophys.J., 615, 209.
\bibitem{Hopkins2006}
Hopkins, A. \& Beacon, J. 2006, Astrophys.J., 651, 142.
\bibitem{Kobayashi2013}
Kobayashi, M., Inoue, Y. \& Inoue, A. 2013, Astrophys.J., 763, 3.
\bibitem{Konno2000}
Konno, O. T., K. \& Kojima,Y. 2000, Astron.\& Astrophys., 356, 234.
\bibitem{Kuroda2016}
Kuroda, T., Kotake, K. \& Takiwaki, T. 2016, Astrophys. J. Lett., 829, L14.
\bibitem{Madau2014}
Madau, P. \& Dickinson, M. 2014, Ann. Rev. Astron \& Astrophys., 52, 415.
\bibitem{Mapelli2018}
Mapelli, M. \& Giacobbo, N. 2018, Mon. Not. R. Astron. Soc., 479, 4391.
\bibitem{Murphy2009}
Murphy, J., Ott, C. \& Burrows, A. 2009, Astrophys. J., 707, 1173.
\bibitem{Regimbau2011}
Regimbau, T. 2011, Res. Astron. Astrophys., 11, 369.
\bibitem{Regimbau2007}
Regimbau, T. \& Chauvineau, B. 2007, Class.Quantum Grav., 24, S627.
\bibitem{Regimbau2001a}
Regimbau, T. \& de Freitas Pacheco, J. 2001a, Astron. \& Astrophys., 376, 381.
\bibitem{Regimbau2001b}
Regimbau, T. \& de Freitas Pacheco, J. 2001b, Astron. \& Astrophys., 374, 182.
\bibitem{Regimbau2006a}
Regimbau, T., \& de Freitas Pacheco, J. 2006a, Astron. \& Astrophys., 642, 455.
\bibitem{Regimbau2006b}
Regimbau, T., \& de Freitas Pacheco, J. 2006b, (RP06), Astron. \& Astrophys., 447, 1.
\bibitem{Sotani2016}
Sotani, H., \& Takiwaki, T. 2016, Phys. Rev. D, 94, 044O43.
\end{thebibliography}
\end{document}